\documentclass[guide]{rmaa}
\usepackage{booktabs,graphics,graphicx,keyval,natbib,paralist,psfrag,trig}
\usepackage[latin1]{inputenc}

\title{RADIAL DEPENDENCE OF EXTINCTION IN PARENT GALAXIES OF SUPERNOVAE}
\author{D. Oni{\' c}, B. Arbutina and D. Uro{\v s}evi{\'c}
   \affil{Department of Astronomy, Faculty of Mathematics, University of Belgrade, Serbia}}

 \fulladdresses{

\item D. Oni\'c, B. Arbutina and D. Uro{\v s}evi{\'c}: Department of Astronomy, Faculty of Mathematics,
University of Belgrade, Studentski trg 16, 11000 Belgrade, Serbia
(donic, arbo, dejanu@matf.bg.ac.yu)}

\shortauthor{ONI{\' C}, ARBUTINA \& URO{\v S}EVI{\' C}}
\shorttitle{RADIAL DEPENDENCE OF EXTINCTION IN PARENT GALAXIES OF
SUPERNOVAE}

\ReceivedDate{2007} \AcceptedDate{2008} \SetYear{2008}

\resumen{El problema de la extinción es el asunto más importante
que se ocupará en del proceso de obtener magnitudes absolutas
verdaderas de corazo'n-se derrumba (pelar-sobre incluyendo) las
supernovas (SNe). El modelo plano-paralelo, usado extensamente en
el pasado, fue demostrado para no describir la extinción
adecuadamente. Intentamos aplicar un modelo alternativo que
introduzca la dependencia radial de la extinción en galaxias del
padre de supernovas. Para la extinción calculadora en nuestra
galaxia utilizamos dos diversos métodos y comparamos los
resultados obtenidos. Nuestro análisis se centra sobretodo en una
muestra elegida del pelar-sobre SNe (Ib/c) para el cual
encontramos la magnitud absoluta máxima intrínseca
$\mathrm{M}_{\mathrm{B}}^{0}=-17.80\pm 0.43$.}

\abstract{The problem of extinction is the most important issue to
be dealt with in the process of obtaining true absolute magnitudes
of core-collapse (including stripped-envelope) supernovae (SNe).
The plane-parallel model, widely used in the past, was shown not
to describe extinction adequately. We try to apply an alternative
model which introduces radial dependance of extinction in parent
galaxies of supernovae. For calculating extinction in our Galaxy
we use two different methods and compare the results obtained. Our
analysis is primarily focused on a chosen sample of
stripped-envelope SNe (Ib/c) for which we find intrinsic peak
absolute magnitude $\mathrm{M}_{\mathrm{B}}^{0}=-17.80\pm 0.43$.}


\begin{document}
\maketitle

\section{Introduction}

 As well known, the supernovae (hereafter SNe) type Ia are
widely used by astronomers as distance indicators due to their
small dispersion in peak absolute magnitude  and similar light
curves. Today, they are thought to originate primarily in the
explosion of a Chandrasekhar mass C-O white dwarf in close binary
system, where the secondary star is filling its Roche lobe and
transferring mass through the Lagrange point to the white dwarf
companion. SNe  type II, on the other hand, are a quite
heterogenous class.  They come from stars of initial mass greater
than approximately 8 $\mathcal{M}_{\odot}$, which can have quite
different properties. The situation is still unclear regarding
stripped-envelope SNe (Ib/c). The progenitors of these SNe are
massive stars that have lost most or all of their hydrogen/helium
envelopes, by strong winds such as in Wolf-Rayet stars or through
mass transfer to a companion star in Roche lobe overflow or a
common envelope phase.

 Whatever the exact scenario for stripped-envelope SNe is, there is
some quite unique physics involved in producing these events
characterized by a complete loss of hydrogen/helium. This
"uniqueness" may lead to a smaller dispersion in their
observational properties (e.g. peak brightness), at least in
comparison to more heterogeneous SNe type II. SNe Ib/c thus may be
the second best "standard candles" among supernovae, after SNe Ia.
Our intention is to analyze this possibility. We will primarily
focus on finding the peak absolute magnitude for SNe Ib/c. In
doing so we must be able to eliminate extinction, which in the
case of all core-collapse SNe should be significant.

The problem of extinction is the most important issue to be dealt
with in the process of obtaining true absolute magnitudes of
core-collapse (including stripped-envelope) SNe. The
plane-parallel model which gives absorption dependent on galaxy
inclination, $\mathrm{A}_{g} = \mathrm{A}_{0} \sec i$, widely used
in the past, was shown not to describe extinction adequately
(Cappellaro 1997). We try to apply an alternative model which
introduces radial dependance of extinction (Hatano et al.\@ 1997,
1998).

A certain trend of dimmer SNe with decreasing distance from the
center of a galaxy was already found by Arbutina (2007a,b). In
present more detailed analysis we have increased the number of SNe
in the sample, applied a new model for extinction in the parent
galaxy, and more carefully calculated Galactic extinction.

\section{The model}

Because of their long-lived progenitors, only SNe Ia have been
observed in elliptical galaxies, which are, practically, gas and
dust free.  In spirals SNe Ia can be found in inter-arm regions
and galaxy's halos. Hence, extinction does not influence their
luminosities as much as in the case of core-collapse SNe (Ib/c and
II) which are all observed in spiral  and irregular galaxies which
have significant amounts of dust.

As just mentioned in the previous section, we have chosen to work
with SNe Ib/c because we expect them to be relatively homogenous
class, at least in comparison to SNe II. Ideally, we should
further separate SN Ib from Ic, but the sample is already too
small for doing this, as we shall see. On the other hand, a sample
of stripped-envelope SNe also comprises peculiar SNe Ic (the so
called hypernovae) which can have very different properties, and
thereby were not included in this analysis.

Extinction has a selective nature,  meaning it depends on
wavelength. Shorter wavelengths are more weakened  and therefore
we expect blue magnitudes (B) to be more extincted than, for
example, visual ones (V). Apparent peak B magnitudes for
stripped-envelope SNe that we have focused on have also turned out
to be more frequently available in our primary data source (Asiago
Supernova Catalogue, Barbon et al.\@ 1999).\footnote{See
Richardson et al.\@ (2006) for a somewhat different analysis of
SNe Ib/c peak magnitudes in the V band.} For the absolute peak
blue magnitude we can generally write:
\begin{equation}
\mathrm{M}_{\mathrm{B}}^{0}=\mathrm{m}_{\mathrm{B}}-\mu-\mathrm{A}_{\mathrm{G}}-\mathrm{A}_{\mathrm{g}}=
\mathrm{M}_{\mathrm{B}}-\mathrm{A}_{\mathrm{g}} ,
\end{equation}
where $\mathrm{m}_{\mathrm{B}}$ is apparent magnitude, $\mu =5\log
(d/ \mathrm{Mpc})$ is distance modulus, $\mathrm{A}_{\mathrm{G}}$
and $\mathrm{A}_{\mathrm{g}}$ are Galactic extinction and
extinction in parent galaxy, respectively.

Bearing in mind the short life of their progenitors, we may assume
that the stripped-envelope SNe are practically in the galactic
plane ($Z = 0$). The radial position of supernova in a galaxy is
then (see Figure 1):
\begin{eqnarray}
\lefteqn{r^{2}=d^{2}(x'^{2}+y'^{2}\sec^{2}i)={} }
                                                    \nonumber\\
&& {}{d^{2}(x^{2}+y^{2})(\cos^{2}(\arctan(\frac{y}{x})+\Pi-90\arcdeg)+{}}  \nonumber\\
&& {}{+\sin^{2}(\arctan(\frac{y}{x})+\Pi-90\arcdeg)\sec^{2}i),}
\end{eqnarray}
where $x$ and $y$ give SN offset from the center of the galaxy in
radians, $\Pi$ is the position angle of the major axis and $d$ is
the distance to the galaxy. If not given, offsets can be
calculated from right ascension and declination;
$x\approx(\alpha_{\mathrm{SN}}-\alpha_{\mathrm{g}})\cos\delta_{\mathrm{g}}$
and $y\approx(\delta_{\mathrm{SN}}-\delta_{\mathrm{g}})$.

\begin{figure}
  \includegraphics[bb=177 288 435 504,width=80mm,keepaspectratio]{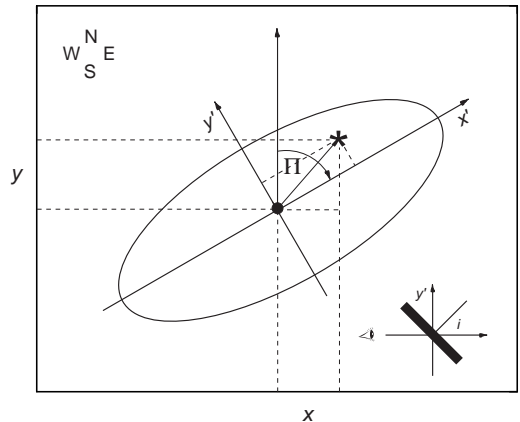}
   \caption{The figure shows how radial position of a supernova in its parent galaxy was calculated.}
\end{figure}

If we assume that
\begin{equation}
\mathrm{A}_{\mathrm{g}}=\mathrm{A}_{0}e^{-a r/R},
\end{equation}
 then
\begin{equation}
\mathrm{M}_{\mathrm{B}}=\mathrm{M}_{\mathrm{B}}^{0}+\mathrm{A}_{0}e^{-a
r/R}.
\end{equation}
 $R$ is the radius of a galaxy, and $\mathrm{A}_{0}$ and $a$
parameters which can be obtained from the fit.

Extinction has been already assumed to follow an exponential law
by Hatano et al.\@ (1997). It is known that dust density in our
Galaxy and M31 actually does not peak at the center (nor in bulge)
but well out in the disk, in a molecular ring where most of the
current star formation is taking place (see Figure 9 in Sodroski
et al.\@ 1997, Hatano et al.\@ 1998). This shortage of the model
is not that important for two reasons: (i) there is a selection
effect against seeing SNe in already bright bulge (Shaw 1979);
(ii) we do not expect (many) stripped-envelope SNe in old
population rich bulge anyway. We also did not include dependence
of inclination in this preliminary model, which would produce
large errors if we had edge-on galaxies with $i \rightarrow
90\arcdeg$. Finally, spiral galaxies are not azimuthally symmetric
(there are spiral arms, of course) and local environment in which
star explodes can be very different from one case to another, an
effect that we can not account for.

\section{Data}

\subsection{Distances}

When possible, the Cepheid-calibrated distance to the parent
galaxy, or a galaxy in the same group of galaxies, was used. The
distance modulus and its uncertainty were taken directly from the
literature (Richardson et al.\@ 2006, Macri et al.\@ 2001, Thim et
al.\@ 2003). The second choice was the distance given in the
Nearby Galaxy Catalogue (Tully 1988), rescaled from
$H_{0}=75\mathrm{km\, s^{-1}\, Mpc^{-1}}$ to $H_{0}=65\mathrm{km\,
s^{-1}\, Mpc^{-1}}$ for consistency with Arbutina (2007a,b).
Uncertainty of 0.2 mag in the distance modulus was adopted as in
Richardson et al.\@ (2006). The third choice was the distance
calculated from the redshift of the parent galaxy (we have
restricted our analysis to the local universe, $z<0.03$) assuming
$H_{0}=65\mathrm{km\, s^{-1}\, Mpc^{-1}}$. The redshifts corrected
to the Cosmic Microwave Background (CMB) Reference Frame were
taken from the June 2007 version of the NED.\footnote{The
NASA/IPAC Extragalactic Database (NED) is operated by the Jet
Propulsion Laboratory, California Institute of Technology, under
contract with the National Aeronautics and Space Administration
(available at http://nedwww.ipac.caltech.edu/).} The uncertainties
ware calculated assuming a peculiar velocity of $300\mathrm{km\,
s^{-1}}$ as in Richardson et al.\@ (2006).

The distance modulus and its uncertainty for SN 1994I were taken
from Richmond et al.\@ (1996)  (see the references therein) who
adopted the surface brightness fluctuations method,  while for SN
2002ap they were taken from Foley et al.\@ (2003) (and references
therein) who used photometry of the brightest red and blue stars
in the system.

\subsection{Peak Apparent Magnitudes}

For most SNe, peak B magnitude was taken from the June 2007
version of the Asiago Supernova Catalogue (hereafter ASC, Barbon
et al.\@ 1999).\footnote{Available at
http://web.pd.astro.it/supern/} Uncertainty of $0.3$  in apparent
magnitudes listed in the ASC, estimated by Richardson et al.\@
(2002), was adopted. The apparent B magnitude at the time of
maximum and its uncertainty for SN 1994I were taken from Richmond
et al.\@ (1996), for SN 1999ex from Stritzinger et al.\@ (2002)
and for SN 2004aw from Taubenberger et al.\@ (2006). The peak
apparent B magnitude for SN 1990B was estimated from Clocchiatti
et al.\@ (2001) with the highest uncertainty associated.

The peak apparent V magnitude and its uncertainty were taken
directly from Richardson et al.\@ (2006) except for SN 1972R and
SN 2004aw  for which they were taken from Leibundgut et al.\@
(1991) and Taubenberger et al.\@ (2006), respectively.

\subsection{Galactic extinction}

Two different methods for calculation of Galactic extinction were
used.  We thought it would be useful to use values from both
methods and see what effect this  will have on the results (see
Willick 1999).

The first choice were values taken from RC3 catalogue (de
Vaucouleurs et al.\@ 1991) who used the older Burstein and Heiles
maps (hereafter BH), which are based on HI column density and
faint galaxy counts (Burstein \& Heiles 1982). On the basis of the
results of the investigation, it is concluded that a reasonable
estimate of the relative accuracy of the HI/galaxy counts method
reported by BH is $0.01$ mag in $E(\mathrm{B-V})$ or $10\%$ of the
reddening, whichever is larger (Burstein \& Heiles 1982). The
second choice were values for Galactic extinction by the Schlegel,
Finkbeiner, and Davis maps (hereafter SFD), based on IRAS/DIRBE
measurements of diffuse IR emission (Schlegel et al.\@ 1998), with
values taken from the NED. The reddening estimates from Schlegel
et al.\@ (1998) have an accuracy of $16\%$.

Galactic extinction
$\mathrm{A}^{\mathrm{SFD}}_{\mathrm{G}}\mathrm{(V)}$ and its
uncertainty were taken directly from Richardson et al.\@ (2006)
except for SN 1972R and SN 2004aw for which they were taken from
NED.

\subsection{Other data}

The position angle of the major axis of the parent galaxy
(measured North Eastwards) was taken from ASC, except for SN
1983N, SN 1984I, SN 1991N and SN 1998bw  for which it was taken
from the July 2007 version of the SAI supernova
catalogue.\footnote{Sternberg Astronomical Institute Supernova
Catalogue (SAI) by D.Yu. Tsvetkov, N.N. Pavlyuk, O.S. Bartunov and
Yu.P. Pskovskii, Sternberg Astronomical Institute, Moscow
University, Moscow, Russia (available at
http://www.sai.msu.su/sn/sncat/).} The SN type, the parent galaxy
type, the inclination of the polar axis with respect to the line
of sight in degrees (0$\arcdeg$ for face on systems), the SN
offset from the galaxy nucleus in arcseconds, and decimal
logarithm of the apparent isophotal diameter in units of $0.1$
arcminutes, were taken from ASC.

 There is some discrepancy regarding the type of SN 1972R (Ib in
ASC, Ia in Patat et al.\@ 1997, Ib? in Leibundgut et al.\@ 1991,
Ipec in SAI supernova catalogue) and SN 1999ex (Ib/c in ASC, Ib in
Richardson et al.\@ 2006). We have marked these as Ib? and Ib/c,
respectivley.

\section{Analysis and results}

Table 1 gives a sample of SNe Ib/c with the known peak apparent B
magnitudes and the calculated SN radial positions. Table 2 gives
redshift and distance modulus for each supernova in the sample.
Table 3 and 4 give absolute B magnitudes at maximum light,
uncorrected for the parent galaxy extinction, for the same sample
of SNe, with BH and SFD model used for calculation of Galactic
extinction, respectively. Table 5 gives peak V absolute
magnitudes, uncorrected for the parent galaxy extinction, with
only SFD method for calculation of Galactic extinction used.

We see in Figure 2 and 4 that there is a certain trend of dimmer
SNe with decreasing distance from the center of a galaxy which can
be attributed to extinction. It can also be seen in Figure 2 that
SNe Ic show larger dispersion then SNe Ib.

Peculiar SNe Ic, also known as hypernovae, show very large
dispersion (see Figure 2), they may have very different properties
and are thereby excluded from the fit. SN 2004aw appears to be a
link between a normal Type Ic supernova like SN 1994I and the
group of broad-lined SNe Ic like 1998bw and 2002ap (Taubenberger
et al.\@ 2006). NGC 3997, host of SN 2004aw, could be a merger
system of two spiral galaxies showing tidally deformed spiral arms
(Taubenberger et al.\@ 2006), which may explain the SN position.
Another supernova that was excluded from the fit is SN 1990B. It
shows some kind of anomalous extinction. The reasons may lie in a
specific environment. Unusually high value
$\mathrm{M}_{\mathrm{B/V}}$ could be due to the high value of the
host galaxy extinction (as estimated in Richardson et al.\@
2006).\footnote{There are few SN Ib/c from ASC excluded from this
analysis; SN 1954A is located in an irregular galaxy, SN 1966J was
shown to be SN Ia (Casebeer et al.\@ 2000), SN 2001B was excluded
because of the unknown peak B luminosity, SN 2006tq because of
some other unknown properties.}

\begin{figure}[h!]
  \includegraphics[bb=0 0 318 244,width=80mm,keepaspectratio]{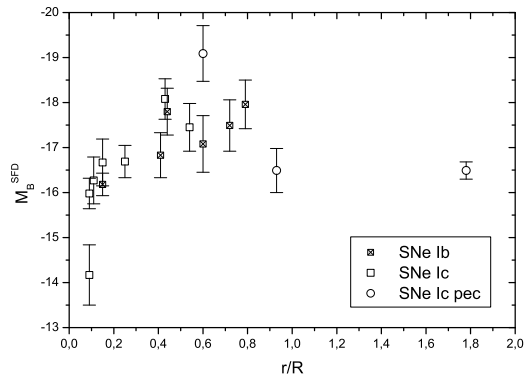}
  \caption{The SN B absolute magnitudes, uncorrected for the parent galaxy extinction, are plotted against the relative radial positions of a SN in the galaxy.
The SFD method for calculation of Galactic extinction is used.}
\end{figure}

In Figure 3 The SN V absolute magnitudes, uncorrected for the
parent galaxy extinction, are plotted against relative radial
position of a SN in the galaxy, with SFD method for calculation of
Galactic extinction used. Contrary to the case of B magnitudes, a
significant trend of dimmer SNe with decreasing radius cannot be
observed here. This is understandable because the extinction
should have weaker influence on V magnitudes than B magnitudes.

\begin{figure}[h!]
  \includegraphics[bb=0 0 318 244,width=80mm,keepaspectratio]{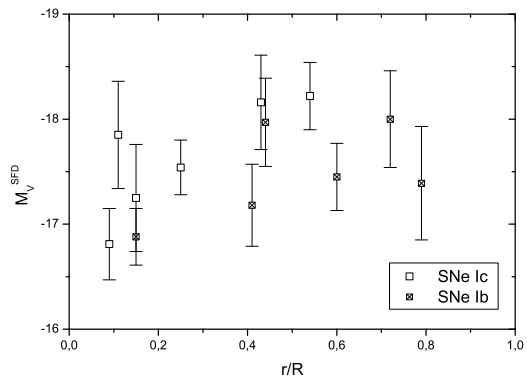}
  \caption{The SN V absolute magnitudes, uncorrected for the parent galaxy extinction, are plotted against relative radial
positions of a SN in the galaxy. The SFD method for the
calculation of Galactic extinction is used. {\bf SNe Ic pec} are
excluded, as well as SN 1990B.}
\end{figure}

 If we assume that extinction is negligible when $r/R \rightarrow
\infty$ a fit to data can give us intrinsic absolute magnitude for
SNe Ib/c (see Figure 4). These are:
$\mathrm{M}_{\mathrm{B}}^{0}=-17.86\pm 0.46$, for BH method used
and $\mathrm{M}_{\mathrm{B}}^{0}=-17.74\pm 0.40$, for SFD method
used. Figure 5 shows the effect of BH/SFD method used in
calculation of Galactic extinction on the assumed extinction trend
and resulting peak absolute magnitude. As we can see the effect is
not large. We will thereby adopt the mean value:
\begin{equation}
\mathrm{M}_{\mathrm{B}}^{0}=-17.80\pm 0.43.
\end{equation}

Curves in Figures 4 and 5 have the form of equations (3) i.e.\@
(4). In Table 6 we give all fit parameters.

\begin{figure}[h!]
  \includegraphics[bb=0 0 318 244,width=80mm,keepaspectratio]{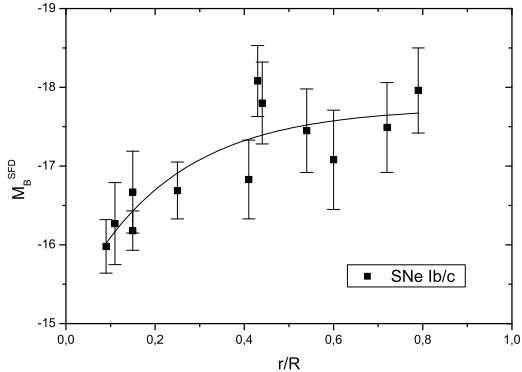}
  \caption{The curve represents the best fit for the data sample of 12 SNe Ib/c.
The SFD method for calculation of Galactic extinction is used. SN
1990B is excluded from the fit.}
\end{figure}

\begin{figure}[h]
  \includegraphics[bb=0 0 339 247,width=80mm,keepaspectratio]{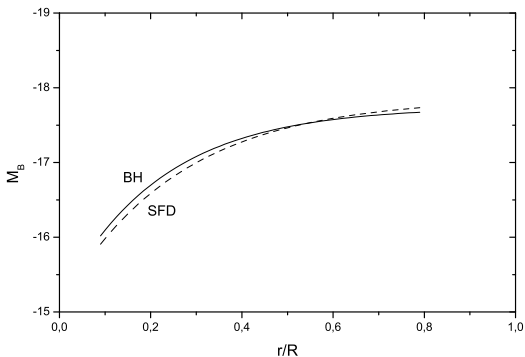}
  \caption{The effect of BH/SFD method used in determining peak absolute magnitude on the results.}
\end{figure}

 It is still possible that there is a substantial range of absolute
magnitudes for Ib/c events. In that case we will have the
Malmquist bias -- only brighter SNe  would be observed at large
distances (high $z$) (see Richardson et al.\@ 2006). Figure 6
shows $\mathrm{M}_{\mathrm{B}}^{\mathrm{SFD}}$ against distance
modulus for analysed Ib/c supernovae. We do not see any distance
dependance, which is quite understandable since we have limited
our analysis to the local universe ($z<0.03$).

\begin{figure}
  \includegraphics[bb=0 0 303 240,width=80mm,keepaspectratio]{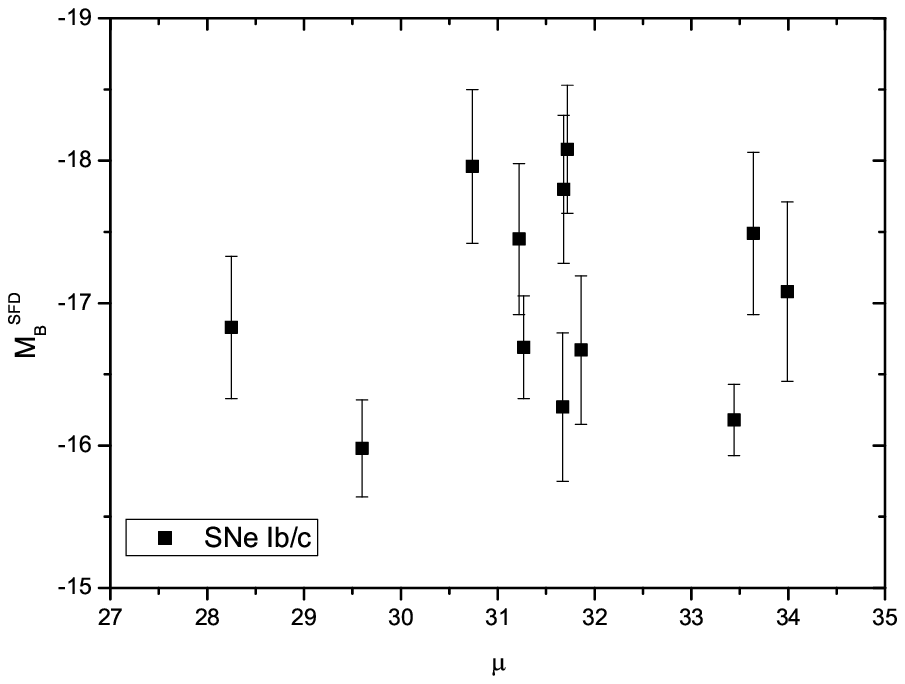}
  \caption{The $\mathrm{M}_{\mathrm{B}}^{\mathrm{SFD}}$ dependance on
distance modulus.}
\end{figure}

\section{Conclusions}

Measuring the radial dependence of controlled samples of SNe in
galaxies appears to be a promising way to constrain the amount and
the distribution of dust in galaxies (Hatano et al.\@ 1998). As
pointed  out by Hatano et al.\@ (1998) the extinction by dust in
parent galaxies can also affect the observed SNe properties.

In this paper we analysed a sample of SNe Ib/c in neighboring
spiral galaxies.  We have chosen to work with SNe Ib/c since we
expect them to be relatively homogenous class, at least in
comparison to SNe II. B magnitudes were considered because the
extinction has stronger influence on them, as well as because of
the larger quantity of data available (in ASC). A certain trend of
dimmer SNe with decreasing distance from the center of a galaxy
was observed. Such an effect can be attributed to extinction. We
have adopted the mean value for SNe Ib/c intrinsic absolute
magnitude:
\begin{equation} \mathrm{M}_{\mathrm{B}}^{0}=-17.80\pm
0.43,
\end{equation}
obtained from the fit.

We made comparison of the results  obtained by using values for
$\mathrm{A}_\mathrm{G}$ from different methods for Galactic
extinction calculation. The effect is not significant. Both
methods give similar results.

To conclude, analysing the dependence of SN observational
properties on radial position in their parent galaxies may hold
some promise. Nevertheless, more data are needed to reach stronger
conclusions.

\vspace{8mm}

The authors would like to thank the referee Prof. David Branch for
useful comments. This paper is a part of the projects "Gaseous and
Stellar Components of Galaxies: Interaction and Evolutions" (No.\@
146012) and "Physics of Sun and Stars" (No.\@ 146003) supported by
the Ministry of Science of Serbia.

 \begin{Code}
  \begin{table*}
    \caption{Data and radial positions for supernovae from the sample \label{label}}
  \begin{center}
      \begin{tabular}{lllcccccc}\hline\hline
   Supernova&SN&Galaxy&Galaxy&Inclination&Position&Diameter&SN radial&Apparent \\
&type&&type&&angle&$D$ [kpc]&position&magnitude\\
&&&&$i [\arcdeg]$&$\Pi [\arcdeg]$&&$r [\mathrm{kpc}]$&$\mathrm{m}_{\mathrm{B}}$ \\
\hline
SN 1972R&Ib?&NGC 2841&Sb&65&147&32&12.6&$12.85\pm 0.30$\\
SN 1983N&Ib&NGC 5236&SBc&21&45&17&3.5&$11.70\pm 0.30$\\
SN 1984I&Ib&E323-G99&SBcd&25&10&32&11.7&$16.60\pm 0.30$\\
SN 1984L&Ib&NGC 991&SBc&28&60&19&4.1&$14.00\pm 0.30$\\
SN 2000H&Ib&IC 454&SBab&58&140&32&9.5&$17.90\pm 0.30$\\
SN 1999ex&Ib/c&IC 5179&Sbc&61&57&33&2.5&$17.35\pm 0.02$\\
\hline
SN 1962L&Ic&NGC 1073&SBc&25&15&24&6.6&$13.94\pm 0.30$\\
SN 1983I&Ic&NGC 4051&SBbc&35&135&34&7.2&$13.70\pm 0.30$\\
SN 1983V&Ic&NGC 1365&SBb&58&32&57&7.2&$14.67\pm0.30$\\
SN 1987M&Ic&NGC 2715&SBc&74&22&33&2.4&$15.30\pm 0.30$\\
SN 1991N&Ic&NGC 3310&SBbc&19&139&17&0.9&$15.50\pm 0.30$\\
SN 1994I&Ic&NGC 5194&Sbc&48&163&24&1.1&$13.77\pm 0.02$\\
SN 1990B&Ic&NGC 4568&Sbc&65&23&20&0.9&$16.89\pm 0.60$\\
\hline
SN 2004aw&Ic pec&NGC 3997&SBb pec&68&130&38&33.4&$18.06\pm 0.04$\\
SN 1998bw&Ic pec&E184-G82&SB&33&150&9&2.7&$14.09\pm 0.30$\\
SN 2002ap&Ic pec&NGC 628&Sc&24&25&22&10.3&$13.11\pm 0.30$\\
\hline\hline
\end{tabular}
    \end{center}
  \end{table*}
\end{Code}

\begin{Code}
  \begin{table*}
    \caption{SN Redshift and distance modulus \label{label}}
  \begin{center}
      \begin{tabular}{lccc}\hline\hline
   Supernova&Redshift&Distance&Ref.\\
&&modulus\\
&$z$&$\mu$& \\
\hline
SN 1972R&0.002128&$30.74\pm 0.23$&1\\
SN 1983N&0.001711&$28.25\pm 0.15$&2\\
SN 1984I&0.010737&$33.64\pm 0.20$&3\\
SN 1984L&0.005110&$31.68\pm 0.20$&4\\
SN 2000H&0.013159&$33.99\pm 0.17$&3\\
SN 1999ex&0.011415&$33.44\pm 0.22$&3\\
\hline
SN 1962L&0.004030&$31.22\pm 0.20$&4\\
SN 1983I&0.002336&$31.72\pm 0.14$&5\\
SN 1983V&0.005457&$31.27\pm 0.05$&5\\
SN 1987M&0.004466&$31.86\pm 0.20$&4\\
SN 1991N&0.003312&$31.67\pm 0.20$&4\\
SN 1994I&0.001544&$29.60\pm 0.30$&6\\
SN 1990B&0.007522&$30.92\pm 0.05$&5\\
\hline
SN 2004aw&0.015914&$34.46\pm 0.14$&3\\
SN 1998bw&0.008670&$32.93\pm 0.28$&3\\
SN 2002ap&0.002192&$29.30\pm 0.14$&7\\
\hline\hline
\end{tabular}

    \end{center}

\begin{flushleft}

\vspace{3mm} \hspace{3cm}

\small REFERENCES. - (1) Macri et al. 2001; (2) Thim et al. 2003;
(3) NED (References are for redshifts); (4) Tully 1988; (5)
Richardson et al. 2006; (6) Richmond et al. 1996; (7) Foley et al.
2003.
\end{flushleft}

    \end{table*}
\end{Code}

\begin{Code}
  \begin{table*}
    \caption{Absolute magnitude $\mathrm{M}_{\mathrm{B}}^{\mathrm{BH}}$ \label{label}}
  \begin{center}
      \begin{tabular}{lcc}\hline\hline
   Supernova&Galactic&Absolute\\
&absorption&magnitude\tabnotemark{a}\\
&$\mathrm{A}_{\mathrm{G}}^{\mathrm{BH}}(\mathrm{B})$&$\mathrm{M}_{\mathrm{B}}^{\mathrm{BH}}$\\
\hline
SN 1972R&$0.00\pm 0.04$&$-17.89\pm0.57$\\
SN 1983N&$0.15\pm 0.04$&$-16.70\pm 0.49$\\
SN 1984I&$0.45\pm 0.04$&$-17.49\pm 0.54$\\
SN 1984L&$0.00\pm 0.04$&$-17.68\pm 0.54$\\
SN 2000H&$1.44\pm 0.04$&$-17.53\pm 0.51$\\
SN 1999ex&$0.00\pm 0.04$&$-16.09\pm 0.28$\\
\hline
SN 1962L&$0.07\pm 0.04$&$-17.35\pm 0.54$\\
SN 1983I&$0.00\pm 0.04$&$-18.02\pm 0.48$\\
SN 1983V&$0.00\pm 0.04$&$-16.60\pm 0.39$\\
SN 1987M&$0.02\pm 0.04$&$-16.58\pm 0.54$\\
SN 1991N&$0.00\pm 0.04$&$-16.17\pm 0.54$\\
SN 1994I&$0.00\pm 0.04$&$-15.83\pm 0.36$\\
SN 1990B&$0.01\pm 0.04$&$-14.04\pm 0.69$\\
\hline
SN 2004aw&$0.00\pm 0.04$&$-16.40\pm 0.22$\\
SN 1998bw&$0.00\pm 0.04$&$-18.84\pm 0.62$\\
SN 2002ap&$0.13\pm 0.04$&$-16.32\pm 0.48$\\
\hline\hline
\end{tabular}
    \end{center}
      \vspace{3mm} \hspace{3cm}
 $^\mathrm{a}$  $\mathrm{M}_{\mathrm{B}}^{\mathrm{BH}}$ is absolute magnitude uncorrected for extinction in the parent galaxy.

    \end{table*}
\end{Code}

\begin{Code}
  \begin{table*}
    \caption{Absolute magnitude $\mathrm{M}_{\mathrm{B}}^{\mathrm{SFD}}$ \label{label}}
  \begin{center}
      \begin{tabular}{lcc}\hline\hline
   Supernova&Galactic&Absolute\\
&absorption&magnitude\tabnotemark{a}\\
&$\mathrm{A}_{\mathrm{G}}^{\mathrm{SFD}}(\mathrm{B})$&$\mathrm{M}_{\mathrm{B}}^{\mathrm{SFD}}$\\
\hline
SN 1972R&$0.067\pm 0.011$&$-17.96\pm 0.54$\\
SN 1983N&$0.284\pm 0.045$&$-16.83\pm 0.50$\\
SN 1984I&$0.452\pm 0.072$&$-17.49\pm 0.57$\\
SN 1984L&$0.119\pm 0.019$&$-17.80\pm 0.52$\\
SN 2000H&$0.991\pm 0.159$&$-17.08\pm 0.63$\\
SN 1999ex&$0.087\pm 0.014$&$-16.18\pm 0.25$\\
\hline
SN 1962L&$0.169\pm 0.027$&$-17.45\pm 0.53$\\
SN 1983I&$0.056\pm 0.009$&$-18.08\pm 0.45$\\
SN 1983V&$0.088\pm 0.014$&$-16.69\pm 0.36$\\
SN 1987M&$0.110\pm 0.018$&$-16.67\pm 0.52$\\
SN 1991N&$0.097\pm 0.016$&$-16.27\pm 0.52$\\
SN 1994I&$0.150\pm 0.024$&$-15.98\pm 0.34$\\
SN 1990B&$0.141\pm 0.023$&$-14.17\pm 0.67$\\
\hline
SN 2004aw&$0.089\pm 0.014$&$-16.49\pm 0.19$\\
SN 1998bw&$0.253\pm 0.040$&$-19.09\pm 0.62$\\
SN 2002ap&$0.301\pm 0.048$&$-16.49\pm 0.49$\\
\hline\hline
\end{tabular}
    \end{center}
      \vspace{3mm} \hspace{3cm}
 $^\mathrm{a}$  $\mathrm{M}_{\mathrm{B}}^{\mathrm{SFD}}$ is absolute magnitude uncorrected for extinction in the parent galaxy.

    \end{table*}
\end{Code}

\begin{Code}
  \begin{table*}
    \caption{Absolute magnitude $\mathrm{M}_{\mathrm{V}}^{\mathrm{SFD}}$ \label{label}}
  \begin{center}
      \begin{tabular}{lccc}\hline\hline
   Supernova&$\mathrm{m}_{\mathrm{V}}$&$\mathrm{A}_{\mathrm{G}}^{\mathrm{SFD}}(\mathrm{V})$&$\mathrm{M}_{\mathrm{V}}^{\mathrm{SFD}}\tabnotemark{a}$\\
\hline
SN 1972R&$13.4\pm 0.3$&$0.052\pm 0.008$&$-17.39\pm 0.54$\\
SN 1983N&$11.3\pm 0.2$&$0.228\pm 0.037$&$-17.18\pm 0.39$\\
SN 1984I&$15.98\pm 0.20$&$0.344\pm 0.055$&$-18.00\pm 0.46$\\
SN 1984L&$13.8\pm 0.2$&$0.091\pm 0.015$&$-17.97\pm 0.42$\\
SN 2000H&$17.30\pm 0.03$&$0.760\pm 0.122$&$-17.45\pm 0.32$\\
SN 1999ex&$16.63\pm 0.04$&$0.067\pm 0.011$&$-16.88\pm 0.27$\\
\hline
SN 1962L&$13.13\pm 0.10$&$0.130\pm 0.021$&$-18.22\pm 0.32$\\
SN 1983I&$13.6\pm 0.3$&$0.043\pm 0.007$&$-18.16\pm 0.45$\\
SN 1983V&$13.80\pm 0.20$&$0.068\pm 0.011$&$-17.54\pm 0.26$\\
SN 1987M&$14.7\pm 0.3$&$0.085\pm 0.014$&$-17.25\pm 0.51$\\
SN 1991N&$13.9\pm 0.3$&$0.075\pm 0.012$&$-17.85\pm 0.51$\\
SN 1994I&$12.91\pm0.02$&$0.115\pm 0.018$&$-16.81\pm 0.34$\\
SN 1990B&$15.75\pm 0.20$&$0.108\pm 0.017$&$-15.28\pm 0.27$\\
\hline
SN 2004aw&$17.30\pm 0.03$&$0.069\pm 0.011$&$-17.23\pm 0.18$\\
SN 1998bw&$13.75\pm 0.10$&$0.194\pm 0.031$&$-19.37\pm 0.41$\\
SN 2002ap&$12.37\pm 0.04$&$0.161\pm 0.026$&$-17.09\pm 0.21$\\
\hline\hline
\end{tabular}
    \end{center}

     \vspace{3mm} \hspace{3cm}
 $^\mathrm{a}$  $\mathrm{M}_{\mathrm{V}}^{\mathrm{SFD}}$ is absolute magnitude uncorrected for extinction in the parent galaxy.

    \end{table*}
\end{Code}

\begin{Code}
  \begin{table}
    \caption{FIT PARAMETERS \label{label}}
  \begin{center}
      \begin{tabular}{@{\extracolsep{-1.4mm}}lccc@{}}
      \hline\hline
   Method&$\mathrm{M}_{\mathrm{B}}^{0}$&$\mathrm{A}_{0}$&$a$ \\
\hline
BH&$-17.86\pm 0.46$&$2.77\pm 0.51$&$3.89\pm 2.55$ \\
SFD&$-17.74\pm 0.40$&$2.60\pm 0.71$&$4.55\pm 3.38$ \\
\hline\hline
\end{tabular}
    \end{center}
    \end{table}
\end{Code}


\begin{thebibliography}

\bibitem[Arbutina(2007a)]{IJMPD} Arbutina, B.\@ 2007a, IJMPD, 16, 1219
\bibitem[Arbutina(2007b)]{AIPC} Arbutina, B.\@ 2007b, AIP Conf. Proc., 938, 202
\bibitem[Barbon et al.(1999)]{AA} Barbon, R., Boundi, V., Cappellaro, E., \& Turatto, M.\@ 1999, A\&A, 139, 531
\bibitem[Burstein \& Heiles(1982)]{AJ} Burstein D. \& Heiles, C.\@ 1982, AJ, 87, 1165
\bibitem[Cappellaro(1997)]{AA} Cappellaro, E., Turatto, M., Tsvetkov, D. Yu., Bartunov, O. S., Pollas, C., Evans R., \& Humuy, M.\@ 1997, A\&A, 322, 431
\bibitem[Casebeer et al.(2000)]{PASP} Casebeer, G. et al.\@ 2000, PASP, 112, 1433
\bibitem[Clocchiatti et al.(2001)]{ApJ} Clocchiatti, A. et al\@ 2001, ApJ, 553, 886
\bibitem[Foley et al.(2003)]{PASP} Foley, R. J. et al.\@ 2003, PASP, 115, 1220
\bibitem[Hatano et al.(1997)]{ApJ} Hatano, K., Branch, D., Fisher, A. \& Starrfield, S.\@ 1997, MNRAS, 290, 113
\bibitem[Hatano et al.(1998)]{ApJ} Hatano, K., Branch, D., \& Deaton, J.\@ 1998, ApJ, 502, 177
\bibitem[Leibundgut et al.(1991)]{AA} Leibundgut, B. et al.\@ 1991, AA, 89, 537
\bibitem[Macri et al.(2001)]{ApJ} Macri, L. M. et al.\@ 2001, ApJ, 559, 243
\bibitem[Patat et al.(1997)]{AA} Patat, F. et al.\@ 1997, A\&A, 317, 423
\bibitem[Richardson et al.(2006)]{AJ} Richardson, D., Branch, D., \& Baron, E.\@ 2006, AJ, 131, 2233
\bibitem[Richardson et al.(2002)]{AJ} Richardson, D., Branch, D., \& Baron, E.\@ 2002, AJ, 123, 745
\bibitem[Richmond et al.(1996)]{AJ} Richmond, M. W. et al.\@ 1996, AJ, 111, 327
\bibitem[Schlegel et al.(1998)]{ApJ} Schlegel, D. et al.\@ 1998, ApJ, 500, 525
\bibitem[Shaw(1979)]{AA} Shaw, R. L.\@ 1979, A\&A, 76, 188
\bibitem[Sodroski et al.(1997)]{ApJ} Sodroski T. J., Odegard, N., Arendt, R. G., Dwek, E., Weiland, J. L., Hauser, M. G., \& Kelsall, T. 1997, ApJ, 480, 173
\bibitem[Stritzinger et al.(2002)]{AJ} Stritzinger, M. et al.\@ 2002, AJ, 124, 2100
\bibitem[Taubenberger et al.(2006)]{MNRAS} Taubenberger, S. et al.\@ 2006, MNRAS, 371, 1459
\bibitem[Thim et al.(2003)]{ApJ} Thim, F. et al.\@ 2003, ApJ, 590, 256
\bibitem[Tully(1988)]{} Tully, R. B.\@ 1988, Nearby Galaxy Catalogue (Cambridge University
Press)
\bibitem[de Vaucouleurs et al.(1991)]{} de Vaucouleurs, G., de Vaucouleurs, A., Corwin, H.G., Buta, R.J., Paturel G. \& Foque, P.\@ 1991, Third Reference Catalogue of Bright Galaxies (Springer-Verlag, New York)
\bibitem[Willick(1999)]{ApJ} Willick, J. A.\@ 1999, ApJ, 522, 647
\end{thebibliography}
\end{document}